\documentclass[prb, twocolumn, superscriptaddress, floatfix]{revtex4}

\usepackage[draft]{hyperref}
\usepackage{textcomp,amsmath,amsfonts,amssymb}
\usepackage{graphicx}
\usepackage{float}

\begin{document}

\title{Experimental characterization of three-wave mixing in a multimode nonlinear
$\mathbf{KTiOPO_4}$ waveguide}
\author{Micha{\l} Karpi\'{n}ski}
\email{mkarp@fuw.edu.pl}

\author{Czes{\l}aw Radzewicz}
\affiliation{Institute of Experimental Physics, University of Warsaw, ul.\ Ho\.z{a} 69, 00-681 Warsaw, Poland}

\author{Konrad Banaszek}
\affiliation{Institute of Physics, Nicolaus Copernicus University, ul.\ Grudziadzka 5, 87-100 Toru\'{n}, Poland}

\begin{abstract}
We report experimental determination of the phase-matching function for type-II three-wave mixing in a periodically poled $\mathrm{KTiOPO}_4$ waveguide in the 792--815~nm spectral region. The measurement was performed by sum-frequency generation of spectrally tuned fundamental components. Strong dependence of the observed signal on the excited spatial modes in the waveguide has been observed and fully interpreted. These results indicate a route to employ the waveguide for spontaneous parametric down-conversion producing photon pairs in well-defined spatial modes.
\end{abstract}

\maketitle

Photon pairs generated in spontaneous parametric down-conversion find multiple applications in the emerging field of quantum-enhanced technologies \cite{SergCPh03, KurtsAPL06, WeinfAPL08}. Standard sources of photon pairs based on bulk $\chi^{(2)}$ crystals suffer from low brightness and complicated modal structure, which hinders their utilization in practical systems. These features can be improved substantially by realizing the down-conversion process in nonlinear waveguides. The spatial structure of the generated photons is then defined by the waveguide, and the confinement of the interacting fields to a small transverse area throughout the waveguide length enhances the efficiency of the down-conversion process. In addition, waveguide manufacturing can include the technique of quasi-phase matching (QPM), which allows one to employ larger nonlinear coefficients that are otherwise inaccessible at desired wavelengths. All these benefits have been demonstrated in recent experiments using periodically poled potassium titanyl phosphate (PP-KTP) waveguides \cite{BanOL01, URen04, FiorMunrOE07, TrifJOB05}.

Typically available PP-KTP waveguides support several transverse modes in the 800~nm wavelength region, where photon pairs have been generated experimentally. This may have a deleterious effect on the modal purity of the generated photons, if the down-conversion process takes place simultaneously to different combinations of spatial modes. However, the phase matching condition in a waveguide depends on propagation constants, which differ between the modes due to intermodal dispersion \cite{RoelJAP94}. This dependence offers a chance to single out a particular down-conversion process into a well-defined pair of transverse modes, thus delivering spatially pure photons.

In this Letter we present an experimental determination of the phase matching characteristics for a type-II three-wave mixing process in a multimode PP-KTP waveguide. This provides detailed spectral and spatial information that is instrumental in designing the down-conversion process with desired modal properties. The measurement was carried out using the inverse process of sum-frequency (SF) generation with spectrally tuned fundamental beams.

The experimental setup is presented in Fig.~1. The sample was a 4.8~mm long PP-KTP structure (AdvR Inc.), which contained a number of ion-exchanged waveguides with a varying poling period, suitable for type-II three-wave mixing with fundamental beams in the 800~nm wavelength region. A selected waveguide was excited with a beam that combined two orthogonally polarized components. The light source was a home-made Ti:Sapphire femtosecond oscillator (800~nm central wavelength, 25~nm FWHM bandwidth) followed by an optical isolator OI. The laser light with polarization rotated to 45\textdegree\  with respect to the plane of the setup was sent to a Michelson-type interferometer. The input beam was split using a polarizing beam splitter PBS1 into two components, whose spectra were tuned individually with narrowband (approx.\ 0.6~nm FWHM) interference filters IF mounted on rotation stages. The spectral range between 792~nm and 815~nm was covered with two sets of filters, designed for 800~nm and 815~nm. The selected filter bandwidth offered a good compromise between the spectral resolution of the measurement and the level of the sum-frequency signal. In order to extract the combined beams from the interferometer their polarizations were rotated by 90\textdegree\ with quarter-wave plates $\lambda/4$ placed in both arms. The piezo-driven mounts for end mirrors PM were used to correct for the varying deviation of the output beam introduced by rotating the interference filters. The feedback signal to control the tilt of the mirrors was obtained from a 4\% pick-off beam splitter BS1 that sent a beam to two quadrant detectors QD. The pick-off beam was also used to monitor the wavelength of the interferometer output with a spectrometer SP.

The prepared beam was focused at the input face of the waveguide using an aspheric lens AL with 8~mm focal length. The output light was collected by a birefringence free, infinity corrected $50\times$ microscope objective. Dichroic mirrors DM and color filters F were used to separate the SF signal from the fundamental. The intensities of the two fundamental components, separated on a polarizing beam splitter PBS3, were measured using silicon photodiodes PD1 and PD2, while the SF signal was directed to a GaP photodiode PD3 connected to a lock-in amplifier. It was ensured that the detectors collected only light transmitted through the waveguide. Flipper mirrors were used to alternatively direct the output beams to CCD cameras in order to record magnified images of transverse intensity distributions at the exit surface of the waveguide.

\begin{figure}[h]
\includegraphics{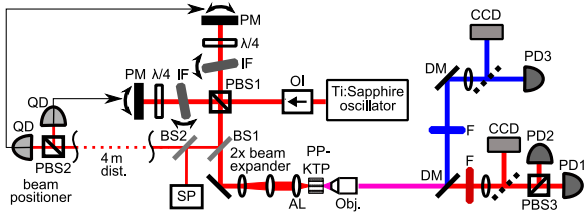}
\caption{Experimental setup. See text for symbol explanation. }
\label{fig:setup}
\end{figure}

\begin{figure}[h]
\includegraphics{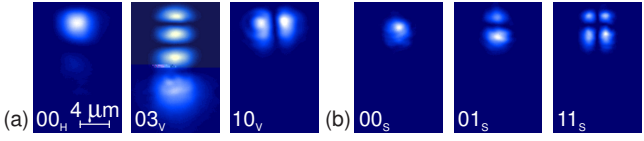}
\caption{Transverse intensity distributions of exemplary modes observed in the waveguide at (a) fundamental and (b) sum-frequency wavelengths, obtained by imaging the output plane of the waveguide onto a CCD camera. }
\label{fig:modes}
\end{figure}

The advantage of the constructed setup was independent control of input coupling for both the components of the fundamental beam by tilting the end mirrors PM. This enabled us to selectively excite specific transverse spatial modes, which was verified with images collected on the CCD camera, as depicted in Fig.~2(a). We label the waveguide modes with two integers $ij$ specifying the number of nodes in directions parallel and perpendicular to the top crystal surface and an index $V$ or $H$ corresponding to the vertical or horizontal polarization of the fundamental component, or $S$ for the sum-frequency. We observed that the SF signal depended critically on the coupling of the fundamental components into specific spatial modes. The SF signal was typically generated in a well defined spatial mode, as shown in Fig.~2(b). These features indicate the important role of intermodal dispersion in the phase matching of three-wave mixing process.

More detailed information about the phase matching can be obtained from a scan of the SF signal intensity as a function of the wavelengths of the fundamental components. In order to aid interpretation of the data, we aimed at exciting well defined spatial modes of the fundamental components throughout the scan. For this purpose, before each scan  the coupling of the fundamental components was adjusted with the end mirrors PM to optimize the intensity of the SF signal at a chosen wavelength. The coupling was then kept fixed during the scan by locking the feedback loop of the piezo drivers to the selected position. The measured sum-frequency intensity was normalized by dividing it by the product of the intensities of the fundamental components measured after the waveguide. This compensated for variable transmission of the interference filters and enabled comparison between scans obtained with different coupling efficiencies.

\begin{figure}[h]
\includegraphics{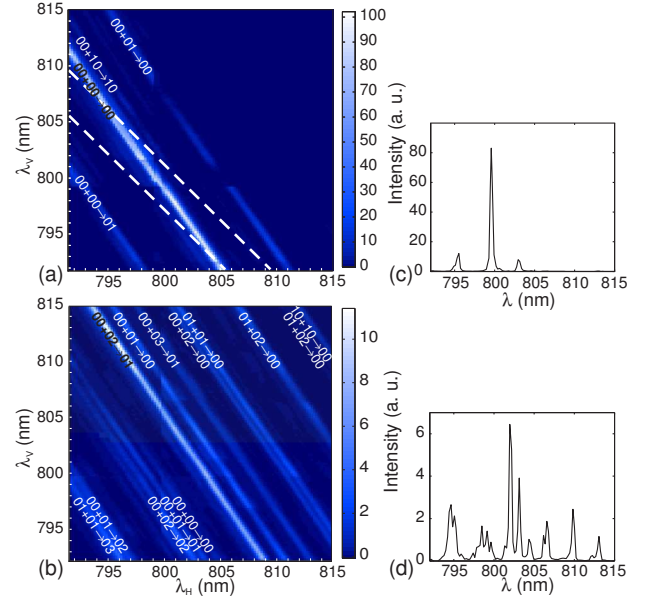}
\caption{The measured normalized sum-frequency intensity vs.\ wavelengths of horizontally and vertically polarized fundamental components (a,b),   scanned with 0.2~nm step, and corresponding one-dimensional cross sections at degenerate wavelengths (c,d). The components were optimized for coupling in (a,c) fundamental $00$ modes and (b,d) a combination of $00_V$ and $02_H$. Each band is labeled with a triplet $ij_V + i'j'_H \rightarrow kl_S$ of interacting spatial modes, identified following the method described in the text. Imperfect excitation of selected spatial modes is seen to generate additional bands. 
The dashed lines in (a) indicate the spectral region of the down-conversion process defined by a pump field centered at $399.8$~nm with $1$~nm FWHM.}
\label{fig:maps}
\end{figure}

The measured maps of the normalized SF intensity are shown in Fig.~3. It is seen that maps are composed of a series of bands. Each band can be associated with a triplet of spatial modes---two fundamental with orthogonal polarizations and one SF---for which the phase matching condition is satisfied. The identification of the relevant modes turned out to be non-trivial due to spectral overlap of the bands and impure excitation of the spatial modes. The starting point of the identification procedure was a series of one-dimensional scans with identical wavelengths for both the fundamental components, similar to those shown in Fig.~3(c,d), carried out for various spatial couplings. This provided unambiguous information on bands that were the brightest or well isolated in the spectral domain.

\begin{figure*}[t!]
\includegraphics{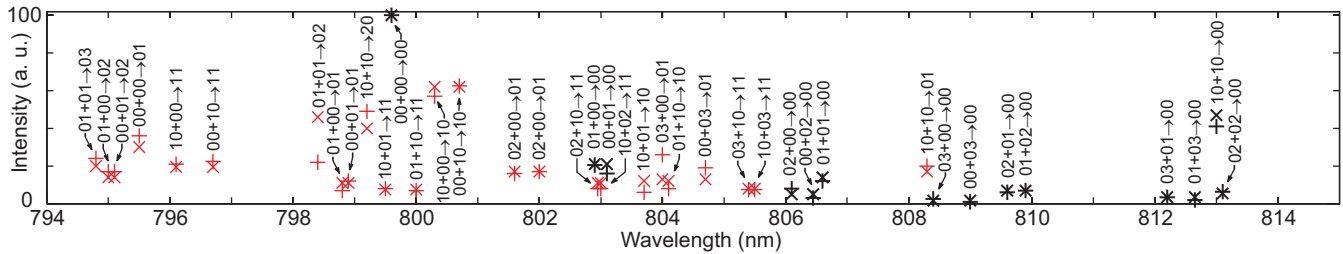}
\caption{Triplets of the interacting spatial modes ordered according to the wavelength of a frequency-degenerate process. Each triplet is represented by a pair of points indicating the measured $(+)$ and calculated $(\times)$ sum-frequency intensity, normalized to $100$ for the $00_V + 00_H \rightarrow 00_S$ band. Triplets involving $00_S$ mode are shown in black, while those involving higher SF modes in red.}
\label{fig:bands}
\end{figure*}

Further identification required a theoretical model.
The phase matching condition for a specific triplet of modes $ij_V + i'j'_H \leftrightarrow kl_S$
in the waveguide is given by the equation:
\begin{equation} \label{Eq:PhMatch}
\omega_V n_{ij}^V(\omega_V) + \omega_H n_{i'j'}^H(\omega_H)
= (\omega_V+\omega_H)n_{{kl}}^S(\omega_V+\omega_H) - \frac{2\pi c}{\Lambda}
\end{equation}
where $\omega_V$ and $\omega_H$ are the frequencies of the two fundamental components, and $n^V_{ij}, n^H_{i'j'}$, and $n^S_{kl}$  are frequency-dependent effective refractive indices of the relevant modes. The parameter $\Lambda$ is the effective grating period of the QPM structure. Refractive indices can be decomposed \cite{RoelJAP94} into the material and geometric contributions according to $n_{ij}^V(\omega) = n^V(\omega) + \Delta n_{ij}^V$, and similarly for $H$ and $S$. Here $n^V(\omega)$ is the refractive index of bulk KTP and the geometric correction $\Delta n_{ij}^V$ is assumed to be constant within the frequency range swept in the experiment. Given two identified triplets that differ only by one of the participating modes, for example $ij_V + i'j'_H \leftrightarrow kl_S$ and $ij_V + i'j'_H \leftrightarrow k'l'_S$, this assumption allows us to determine the difference $\Delta n_{kl}^{S} - \Delta n_{k'l'}^{S}$. Using bulk KTP refractive indices \cite{KatoTakaAO02} and data obtained from one-dimensional scans, we were thus able to calculate relative spectral location of other pairs of bands that involved $\Delta n_{kl}^{S}$ and $\Delta n_{k'l'}^{S}$. These predictions matched closely the locations of bands revealed by one-dimensional scans, with deviations less than $0.2$~nm. The procedure yielded a complete identification of triplets, as indicated in Fig.~3(a,b).

The intensity of the sum-frequency signal is proportional to the squared spatial overlap of the triplet of the modes involved \cite{FiorMunrOE07}. We calculated electric field distributions of the modes using a vector finite-difference modesolver \cite{FallahJLT08}, assuming error function depth profile with refractive index modulation taken from Roelofs \emph{et al.}\cite{RoelJAP94} The geometric parameters of the waveguide were obtained by coupling in white light and analyzing the image of the output face of the waveguide. The theoretically calculated squared overlaps are compared in Fig.~4 with normalized intensities of the sum-frequency signal measured for degenerate wavelengths of $V$ and $H$ components. In this measurement, for each selected band the intensity was maximized by manually adjusting the input coupling. A reasonable agreement between measured and simulated data is seen. Discrepancies can be attributed to impure mode excitation and imperfect numerical modeling.

The experimental data presented in this Letter provide detailed information
on the role of transverse spatial modes in three-wave mixing, which
can be used to engineer the process of spontaneous parametric down-conversion. Suppose that the pump field $S$ is launched into the waveguide in the fundamental $00_S$ mode. Using Fig.~4 it is easy to identify the pairs of modes in which down-converted photons can be produced, and the location of the respective spectral bands in the $(\lambda_V,\lambda_H)$ plane. Thanks to intermodal dispersion, the center of the $00_V + 00_H \leftrightarrow 00_S$ band is shifted from other combinations by more than 3~nm. Therefore the $00_S \rightarrow 00_V + 00_H$ process can be separated by using a long waveguide, for which the bands become sufficiently narrow. Further, the wavelengths of the generated photons must lie within a region defined by the bandwidth of the pump field, as marked in Fig.~3(a) with dashed lines. Because the group velocities for $V$ and $H$ polarized photons are unequal, the slope of the phase matching bands will differ from the slope of the region defined by the pump field. As a result, photons generated in different pairs of spatial modes will be separated spectrally. This should allow one to isolate $00_V+00_H$ pairs by employing spectral filtering only. Note that the spatial purity of the pump field mode is critical in this scheme, as in the vicinity of the $00_S \rightarrow 00_V + 00_H$ band there are other bands involving higher-order $S$ modes. Thus we see that the interplay between spatial and spectral degrees of freedom provides a wealth of tools to engineer the down-conversion process.

We acknowledge helpful discussions with I.\ A.\ Walmsley and C.\ Silberhorn. This work was supported by EU~FP6 project QAP (Contract no.\ 015848) and MNiSW grant no.\ N~N202~1489~33.

\end{document}